\documentclass[reprint]{revtex4-2}
\usepackage{graphicx}
\usepackage{dcolumn}
\usepackage{bm}
\usepackage[mathlines]{lineno}
\usepackage[dvipsnames]{xcolor}
\usepackage{amsmath}
\usepackage{gensymb}
\usepackage[labelfont=bf]{subcaption}
\usepackage{caption}
\usepackage{xr}
\usepackage{setspace}
\usepackage{hyperref}
\usepackage{amsmath}
\hypersetup{
  colorlinks   = true, 
  urlcolor     = blue, 
  linkcolor    = blue, 
  citecolor   = red 
}
\newcommand*{\addFileDependency}[1]{
\typeout{(#1)}
\@addtofilelist{#1}
\IfFileExists{#1}{}{\typeout{No file #1.}}
}\makeatother

\graphicspath{ {./figures/} }
\begin{document}

\preprint{APS/123-QED}

\title{Gate modulation of the hole singlet-triplet qubit frequency in germanium}

\author{John Rooney}
 \affiliation{Physics and Astronomy Department, University of California, Los Angeles}
\author{Zhentao Luo}
\affiliation{Physics and Astronomy Department, University of California, Los Angeles}
\author{Lucas E. A. Stehouwer}
\affiliation{QuTech and Kavli Institute of Nanoscience, Delft University of Technology}%
\author{Giordano Scappucci}
\affiliation{QuTech and Kavli Institute of Nanoscience, Delft University of Technology}%
\author{Menno Veldhorst}
\affiliation{QuTech and Kavli Institute of Nanoscience, Delft University of Technology}%
\author{Hong-Wen Jiang}
\affiliation{Physics and Astronomy Department, University of California, Los Angeles}

\date{\today}

\begin{abstract}
Spin qubits in germanium gate-defined quantum dots have made considerable progress within the last few years, partially due to their strong spin-orbit coupling and site-dependent $g$-tensors. While this characteristic of the $g$-factors removes the need for micromagnets and allows for the possibility of all-electric qubit control, relying on these $g$-tensors necessitates the need to understand their sensitivity to the confinement potential that defines the quantum dots. Here, we demonstrate a $S-T\_$ qubit whose frequency is a strong function of the voltage applied to the barrier gate shared by the quantum dots. We find a g-factor that can be approximately increased by an order of magnitude adjusting the barrier gate voltage only by 12 mV. We attribute the strong dependence to a variable strain profile in our device. This work not only reinforces previous findings that site-dependent $g$-tensors in germanium can be utilized for qubit manipulation, but reveals the sensitivity and tunability these $g$-tensors have to the electrostatic confinement of the quantum dot.   
\\
\end{abstract}

\maketitle

\section{\label{sec:introduction} Background and Objectives}

Utilizing hole spin states in strained germanium (Ge/SiGe) gate-defined quantum dots for qubit operation has developed rapidly over the past several years, as groups have demonstrated fast two-qubit logic \cite{Hendrickx2020}, singlet-triplet encodings \cite{Jirovec2021}. and a four-qubit quantum processor \cite{Hendrickx2021}. The success of these experiments can partly be attributed to the various advantages of holes for spin qubit encoding \cite{Scappucci2021}. In stark contrast to electrons, the two topmost valence bands in Ge are well separated in energy due to strain and 2D confinement. The light effective mass (0.054m$_e$ \cite{Lodari2021}) of holes in the topmost band and the absence of valley degeneracy allows us to easily access the two highest hole states for spin encoding. Furthermore, Ge hole spin coherence times benefit from their weak hyperfine interaction with surrounding nuclear spins. Finally, because of their strong spin-orbit coupling and site-dependent g-tensors, Ge hole quantum dots do not require the fabrication of micromagnets, advancing their potential for scalability and integration into current industrial semiconductor facilities \cite{Scappucci2021}.  

Most double quantum dot singlet-triplet qubit studies have focused on encodings between the singlet $|S\rangle$ and unpolarized triplet $|T_0\rangle$ states \cite{Jirovec2021,Wu2014,Petta2005,Maune2012}. In this work, we detail the dynamics of the $S-T\_$ subspace, which has been less studied thus far \cite{Jirovec2022,WangScappucci2022,Wang2023}. Furthermore, we explore the tunability of the hole $g$-tensors by varying the electrostatic potential generated by the barrier gate bridging the two quantum dots. As quantum computation with Ge hole spins critically depends on the $g$-tensor, the ability to manipulate the $g$-tensor becomes a valuable asset for a spin qubit encoded in this system.   

\section{\label{sec:results} Results}

\begin{figure}
    \begin{subfigure}[t]{0.42\columnwidth}
        \caption{\label{fig: device SEM}}
        \includegraphics[width=\linewidth]{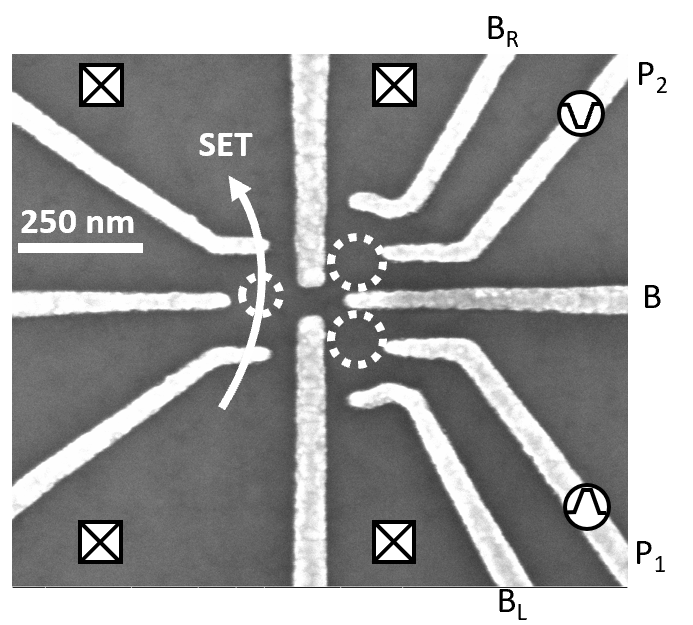}  
    \end{subfigure}
    \begin{subfigure}[t]{0.45\columnwidth}
         \caption{\label{fig: heterostructure}}
         \includegraphics[width=\linewidth]{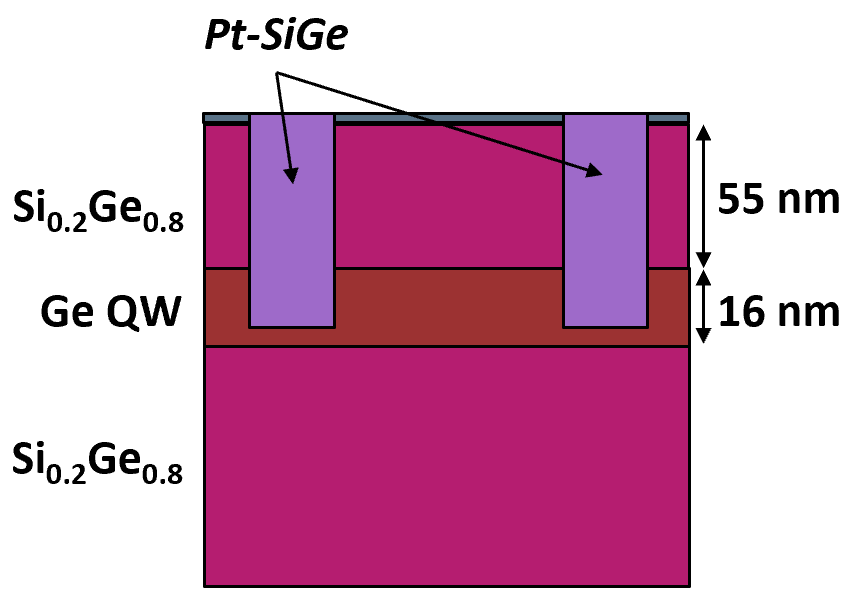}
     \end{subfigure} 
    \begin{subfigure}[t]{0.45\columnwidth}
         \caption{\label{fig: large stability diagram}}
         \includegraphics[width=\linewidth]{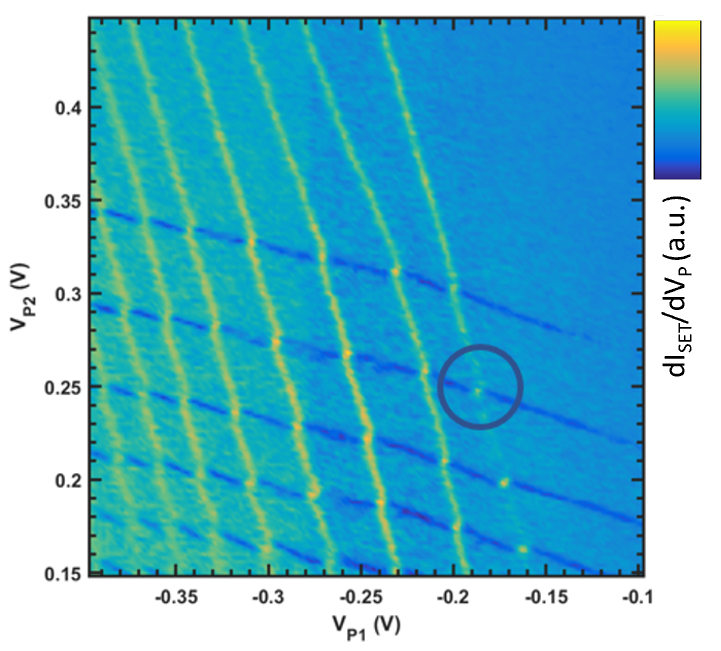}
     \end{subfigure}
    \begin{subfigure}[t]{0.46\columnwidth}
        \caption{\label{fig: small stability diagram}}
        \includegraphics[width=\linewidth]{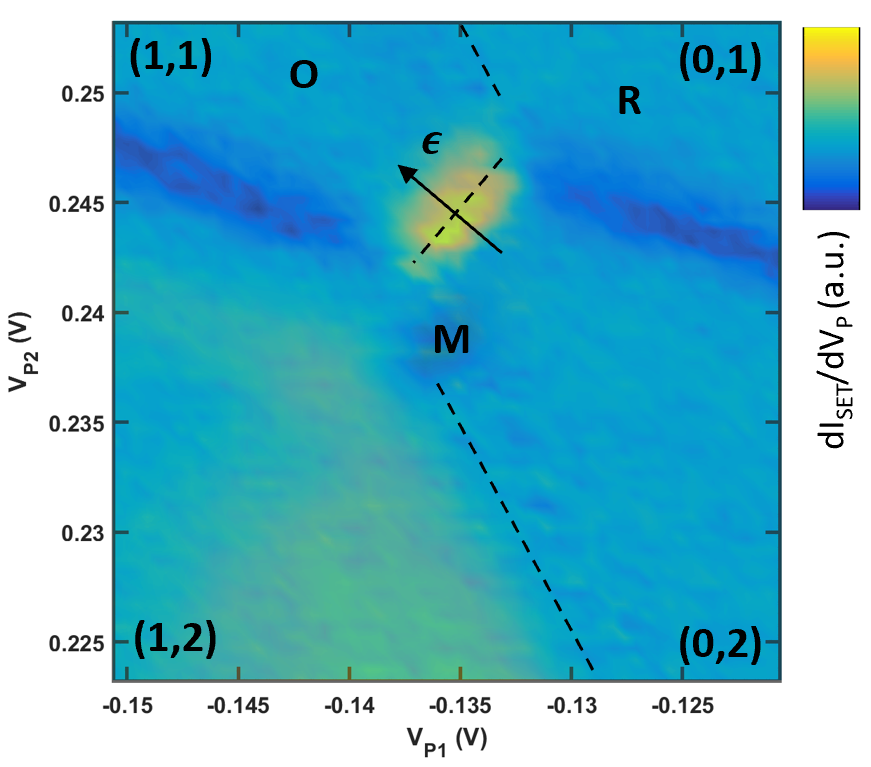}
    \end{subfigure}
\caption{\label{fig: device}{\textbf{(a)} SEM image of lithographically defined gates identical to the device used in this study. \textbf{(b)} Heterostructure of the device showing the Ge quantum well packed between two Ge rich SiGe layers. \textbf{(c)} A typical stability diagram with the circle highlighting the (1,1)-(0,2) anticrossing. \textbf{(d)} All experiments were completed at the (1,1)-(0,2) anticrossing, where ($n$,$m$) denotes the hole occupation for each dot. Point R was used to reset the DQD, M for measurement and initialization, and O for coherent operation between the singlet and triplet states.}}
\end{figure}

A scanning electron microscope image of the device studied is shown in Fig. \ref{fig: device SEM} along with the Ge/SiGe heterostructure in Fig. \ref{fig: heterostructure}. The strained Ge quantum well is 16 nm in width and located 55 nm below the surface. For more details regarding the heterostructure, see \cite{Lodari2021}. A two-dimensional hole gas is first created in the Ge well by applying a negative voltage to a global top gate situated above the heterostructure. The double quantum dot (DQD) is then formed underneath plungers P$_1$ and P$_2$ by applying appropriate voltages to the neighboring barrier gates, where the middle barrier voltage $V_B$ controls the coupling between the two dots. Varying the plunger voltages controls the chemical potential of each dot, allowing us to reach the few-hole regime (Fig. \ref{fig: large stability diagram}), where all experiments were performed at the (1,1)-(0,2) anticrossing (Fig. \ref{fig: small stability diagram}). The hole occupation of both dots was detected by the nearby SET (left half of the device) labeled in Fig. \ref{fig: device SEM}. For convenience in describing this DQD system, we define the relative energy of the two quantum dots as the detuning $\epsilon=\alpha_2 V_{P2} - \alpha_1 V_{P1}$, where $\alpha_i$ converts the voltage applied to P$_i$ to the change in the energy level of dot $i$. Fig. \ref{fig: small stability diagram} illustrates the detuning axis on the stability diagram, where $\epsilon=0$ at the (1,1)-(0,2) boundary. 

\begin{figure}[h]
    \begin{subfigure}[t]{0.47\columnwidth}
         \caption{\label{fig: STminus oscillations}}
         \includegraphics[width=\linewidth]{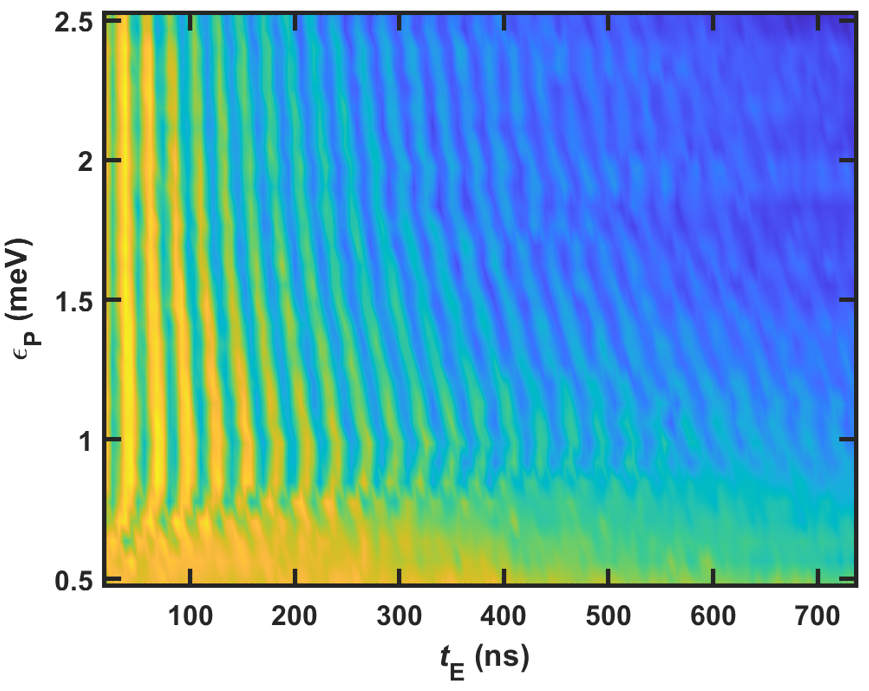}
     \end{subfigure}
    \begin{subfigure}[t]{0.05\linewidth}
    \hspace{-0.46cm}
        \raisebox{-2.9cm}{%
        \includegraphics[scale=0.5]{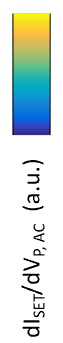}}
    \end{subfigure}
    \begin{subfigure}[t]{0.45\columnwidth}
         \caption{\label{fig: STminus oscillations FFT}}
         \includegraphics[width=\linewidth]{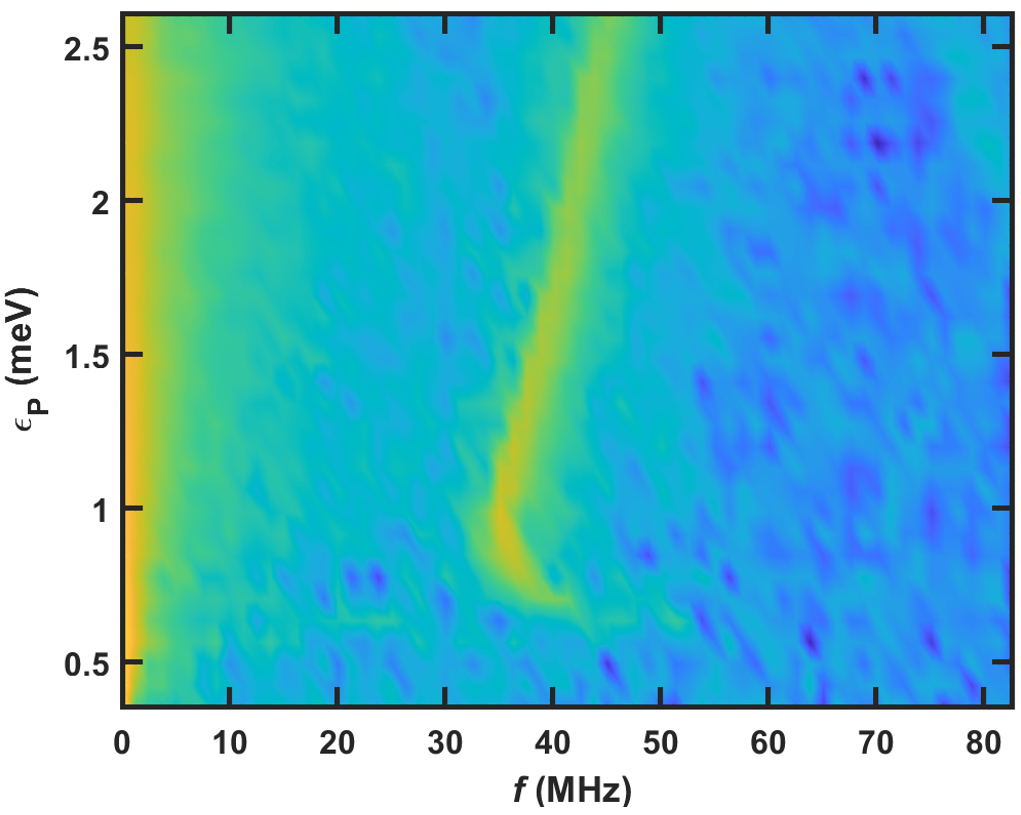}
     \end{subfigure}     
    \begin{subfigure}[t]{0.45\columnwidth}
        \caption{\label{fig: energy dispersion}}
        \includegraphics[width=\linewidth]{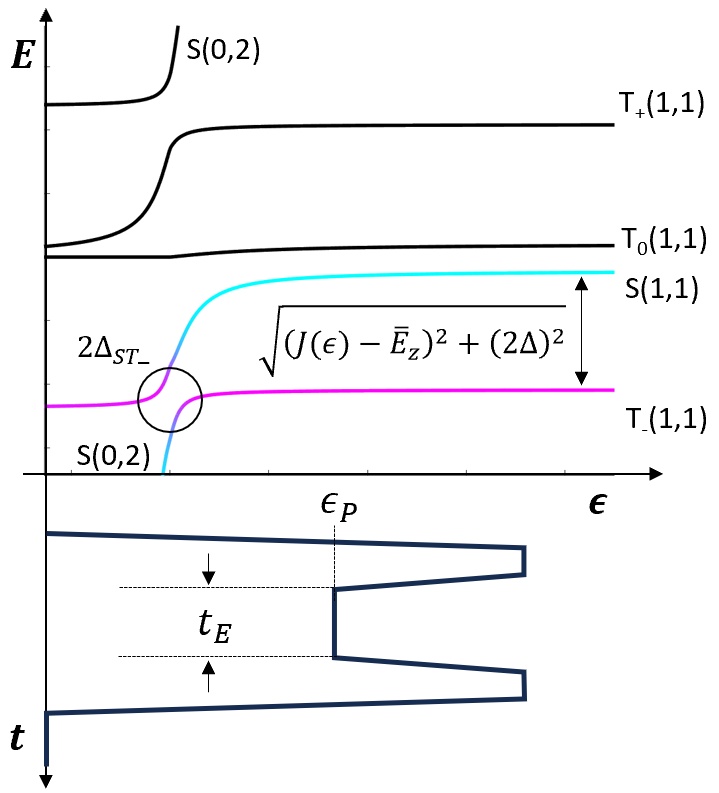}
     \end{subfigure}  
     \hspace{0.05\linewidth}
    \begin{subfigure}[t]{0.45\columnwidth}
         \caption{\label{fig: bloch sphere}}
         \includegraphics[width=\linewidth]{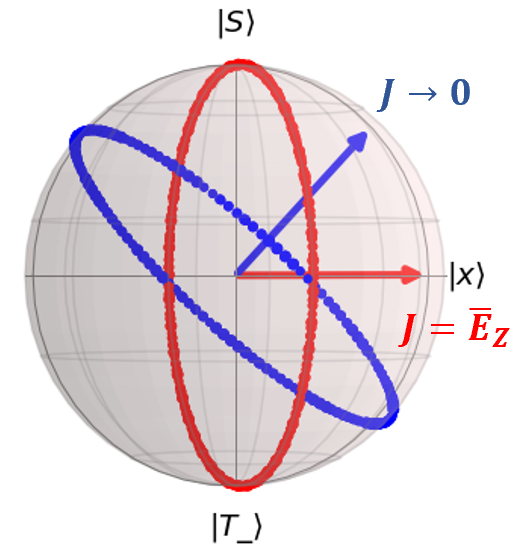}
     \end{subfigure}  
\caption{\label{fig: coherent oscillations} {\textbf{(a)} The SET signal as a function of the detuning and evolution time, illustrating coherent oscillations between $|S\rangle$ and $|T\_\rangle$. The chevron pattern located near 1 meV arises from the $S-T\_$ anticrossing defined by the energy splitting $\Delta_{ST\_}$, while oscillations at large detunings are controlled by the average Zeeman energy of the two dots: $\overline{E}_z$. \textbf{(b)} Fourier transform of the coherent oscillations in (a), illustrating the $S-T\_$ energy splitting as a function of detuning. \textbf{(c)} Energy levels (not to scale) of the singlet and triplets as a function of detuning. The Ramsey pulse used is shown below as a function of time and detuning. \textbf{(d)} Bloch sphere depicting the two rotation axes for the $S-T\_$ subspace. When $\epsilon_P$ is at the $S-T\_$ anticrossing, the system undergoes X rotations (red axis). For large detunings, a combination of X and Z rotations are performed (blue).}}
\end{figure}

When the system passes the $\epsilon = 0$ detuning line into the (1,1) charge configuration, the (0,2) singlet state hybridizes with the (1,1) singlet due to the tunnel coupling between the quantum dots: $|S\rangle = \sin{\left(\Omega/2\right)}|S_{02}\rangle- \cos{\left(\Omega/2\right)}|S_{11}\rangle$. Here, $\Omega=\arctan{\left(\frac{2\sqrt{2}t_c}{\epsilon}\right)}$ is the mixing angle between the two singlet states. In addition to $|S\rangle$, Fig. \ref{fig: energy dispersion} depicts the three triplet states that compose the four lowest energy levels in the (1,1) charge configuration. A simple block magnet situated near the device’s PCB provided the field necessary to lift the degeneracy of the three triplet states, generating an estimated fixed global out-of-plane field of 1.2 mT and in-plane field of 4.4 mT measured at the device's position ($\mathbf{|B|} = B = 4.6$ mT points $\theta = 15\degree$ out of the x-y plane). This tilted field differs from previous qubit experiments on this heterostructure where $B$ was completely in-plane, allowing for a unique perspective into the hole spin states \cite{Hendrickx2021,Wang2023}. Importantly, this magnetic field splits the polarized triplet $|T\_\rangle$ from $|T_0\rangle$ by the average Zeeman energy of the quantum dots $\overline{E}_z$.    
 
Beginning at M in Fig. \ref{fig: small stability diagram}, the system is first initialized into the (0,2) singlet state. A voltage pulse was then applied to P$_1$ and P$_2$ to quickly separate the holes and create a small admixture between the (1,1) singlet $|S\rangle = \frac{1}{\sqrt{2}}(|\uparrow\downarrow\rangle - |\downarrow\uparrow\rangle)$ and polarized triplet $|T\_\rangle = |\downarrow \downarrow\rangle$ states. Once the holes were separated, the system was pulsed to various operation detunings $\epsilon_P$ and allowed to evolve for a time $t_E$ between $|S\rangle$ and $|T\_\rangle$ (Fig. \ref{fig: STminus oscillations} and \ref{fig: energy dispersion}). The qubit frequency (Fig. \ref{fig: STminus oscillations FFT}) is given by the energy difference between these two states at the operation detuning: $hf=\Delta E_{ST\_}$, which plateaus to roughly $\overline{E}_z$ for large detunings. 

For smaller operation detunings, the energy splitting reaches a minimum at the $S-T\_$ anticrossing, where it approximately equals $2\Delta_{ST\_}$. We define $\Delta_{ST\_}$ as the coupling between $|S\rangle$ and $|T\_\rangle$ at the $S-T\_$ anticrossing. By varying the operation detuning $\epsilon_P$ from 0.5 to 2.5 meV, we sampled the energy splitting between the two lowest states for both regimes. The existence of this minimum leads to the observed chevron pattern at 1 meV in Fig. \ref{fig: STminus oscillations}, which has been seen in previous $S-T\_$ works and absent from studies coherently manipulating the $S-T_0$ states \cite{Jirovec2021,Jirovec2022,Wang2023,Wu2014,WangScappucci2022,Petta2005,Maune2012}.  

To understand these dynamics, we utilize a Hamiltonian describing the $\{|S\rangle,|T\_\rangle\}$ subspace that was derived in Ref. \cite{Mutter2021} and is a reduced form of the full model used in Ref. \cite{Jirovec2022}. To leading order, it takes the following form:
\begin{gather}
\label{eq: H}
    H=
    \begin{pmatrix}
        -J(\epsilon) & \Delta \\
        \Delta & -\overline{E}_z
    \end{pmatrix}.
\end{gather}

\noindent We define the exchange energy $J(\epsilon) = -\frac{\epsilon}{2}+\sqrt{\frac{\epsilon^2}{4}+2t_c^2}$ as the energy difference between $|S\rangle$ and $|T_0\rangle$. The coupling of the $S-T\_$ states ($\Delta$) emerges from two sources: (1) a spin-orbit splitting ($\Delta_{so}$) and (2) an effective Zeeman splitting due to the anisotropy of the $g$-tensors ($g_a$) that is only present when $B$ has non-zero in- and out-of-plane components: $\Delta = |\Delta_{so} \sin{\left(\frac{\Omega}{2}\right)} + g_a \mu_B B \cos{\left(\frac{\Omega}{2}\right)}|$ \cite{Jirovec2022,Mutter2021}. The anisotropy between the in- and out-of-plane $g$-factors of a quantum dot has been previously observed, where the in-plane $g$-factors ($g_\parallel$) were measured to be a few tens to hundreds of times smaller than their out-of-plane counterparts ($g_\perp$) for holes in Ge/SiGe substrates \cite{Jirovec2022,Hendrickx2023}. The $|T\_\rangle$ state splits from $|T_0\rangle$ by the average Zeeman energy, $\overline{E}_z = \overline{g} \mu_B B$, where $\overline{g}$ is the average $g$-factor of the two dots projected onto the axis of $\mathbf{B}$.

With this Hamiltonian, we can solve for the frequency of the $S-T\_$ evolution: $f=\frac{1}{h}\sqrt{(J-\overline{E}_z)^2+(2\Delta)^2}$. At the $S-T\_$ anticrossing, $J=\overline{E}_z$, and $f$ is controlled by $\Delta$, where X rotations are performed around the Bloch sphere (Fig. \ref{fig: bloch sphere}). For large detunings, $J\rightarrow 0$, leaving $f$ to be determined by the average Zeeman energy and $S-T\_$ coupling, and the qubit rotates near the z axis. The larger the ratio $\frac{\overline{E}_z}{\Delta}$ becomes, the closer this axis aligns with the z direction. We note that with control over the orientation of the magnetic field, it is possible for $\Delta \rightarrow 0$ at specific detunings, resulting in perfect Z rotations \cite{Mutter2021}.    

After manipulation, the separated holes were reunited in the (0,2) charge configuration at M for spin readout of the final state using Pauli spin blockade. The system is then reset at R before repeating the cycle again. For a detailed explanation of each step of the pulse, see section S\ref{sec: AWG sequence} of the supplementary material. 

\begin{figure}[h]
    \begin{subfigure}[t]{0.48\columnwidth}
        \caption{\label{fig: T2* plot}}
        \includegraphics[width=\linewidth]{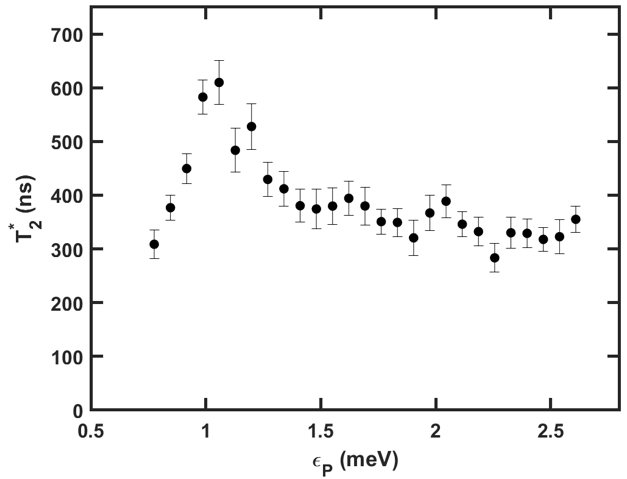}  
    \end{subfigure}
    \begin{subfigure}[t]{0.48\columnwidth}
        \caption{\label{fig: T1 plot}}
        \includegraphics[width=\linewidth]{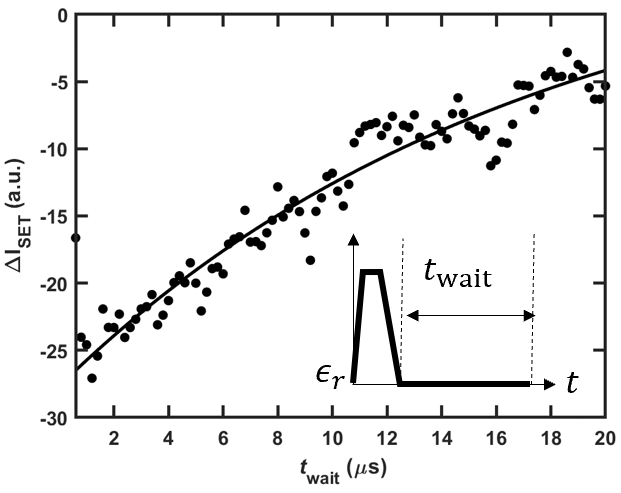}
    \end{subfigure}
\caption{\label{fig: decoherence}{Dephasing and relaxation measurements \textbf{(a)} $T_2^*$ as a function of detuning. Linecuts in Fig. \ref{fig: STminus oscillations} are fit to the curve $P=Ae^{-(t/T_2^*)^2}\cos{(\omega t+B)}+Ce^{-t/D}+E$, where the dephasing time $T_2^*$ is extracted and plotted in (a). Error bars equal one standard deviation of the uncertainty in $T_2^*$ from this fit. This decoherence can be understood as a contribution from two noise terms with the simple model shown in Eqn. \ref{eq: T2*}. From this model, we estimate $\delta \Delta_\text{rms}=0.8$ and $\delta \overline{E}_{z,\text{rms}}=3$ neV. \textbf{(b)} A $T_1$ measurement where the change in SET current is recorded as a function of wait time at the measurement point M. We extract $T_1$ from the fit $P=A e^{\left(-t/T_1 \right)}+B$ (solid line) and calculate $T_1 = 17.2\pm3.2$ $\mu$s. Inset: the pulse used to observe this decay.}}
\end{figure}

We first analyzed the dephasing and relaxation of this qubit by measuring $T_2^*$ and $T_1$. For each $\epsilon_P$ in Fig. \ref{fig: STminus oscillations}, the $S-T\_$ evolution was fit to a Gaussian damped sinusoid $P=Ae^{-(t/T_2^*)^2}\cos{(\omega t+B)}+Ce^{-t/D}+E$, where $T_2^*$ is the inhomogeneous dephasing time. For example traces and details relating to this fit, see section S\ref{sec: T2*}. After extracting $T_2^*$ as a function of $\epsilon_P$ (Fig. \ref{fig: T2* plot}), a clear dependence on the pulse height is seen. This behavior can be understood with a simple model describing the influence of charge and magnetic noise on the fluctuations in the energy difference between the two states \cite{Wu2014,Jirovec2021}: 
\begin{equation}
\label{eq: T2*}
\sqrt{2}\hbar T_2^{*-1} =\sqrt{{\langle \delta E^2\rangle} }. 
\end{equation}

At the $S-T\_$ anticrossing where the qubit frequency reaches a minimum, the system is insensitive to first-order to fluctuations in $\epsilon$ due to charge noise. This protection leads to the maximum in $T_2^*$ seen at 1 meV in Fig. \ref{fig: T2* plot}. However, the qubit is still susceptible to electrical noise affecting the dot $g$-factors and tunnel coupling as well as magnetic noise afflicting $B$. We can estimate the magnitude of this noise combination from Eqn. \ref{eq: T2*} using the fact that $J = \overline{E}_z$ at this detuning. Under this condition $\delta E = 2\delta \Delta_\text{rms}$, where we define $\delta \Delta_\text{rms}$ to include the noise sources pertinent to $t_c$, $g_a$, and $B$, leading to  

\begin{equation*}
    \delta \Delta_\text{rms} \approx \frac{\sqrt{2} \hbar}{2} (T_2^* = 600 \text{ ns})^{-1} = 0.8 \text{ neV}. 
\end{equation*}

For large operation detunings, the energy separation between the $S-T\_$ states reaches a parallel regime (Fig. \ref{fig: STminus oscillations FFT}), which diminishes the charge noise contribution to $\delta E$. In this regime, $\Delta E_{ST\_}$ approximately equals $\overline{E}_z$, where the combined electrical and magnetic "Zeeman" noise affecting $\overline{g}$ and $B$ limits $T_2^*$. We estimate this parameter using Eqn. \ref{eq: T2*} again:   

\begin{align*}
    \delta E &\approx \delta \overline{E}_{z,\text{rms}}, \\
    \delta \overline{E}_{z,\text{rms}} &\approx \sqrt{2} \hbar (T_2^* = 317 \text{ ns})^{-1} = 3 \text{ neV}.
\end{align*}

The system's $T_1$ spin relaxation time was measured by varying the wait time at the readout position $\epsilon_r$ (Fig. \ref{fig: T1 plot}). For these measurements, the system was allowed to completely dephase at the operation detuning $\epsilon_P$ before being pulsed back to the readout window for a variable amount of time (see inset of Fig. \ref{fig: T1 plot}). We fit the resulting exponential decaying curve shown in Fig. \ref{fig: T1 plot} to $P=Ae^{-(t/T_1)}+B$ and find $T_1$ to be $17.2\pm 3.2$ $\mu$s, which is comparable to experiments done in single hole \cite{Hendrickx2020} and $S-T_0$ qubits \cite{Jirovec2021}. However, this $T_1$ can still be improved, as single hole spin relaxation times as long as 32 ms with $B=0.67$ T have been measured using tighter dot confinements and limiting the dot-reservoir coupling \cite{Lawrie2020}. 

\begin{figure*}
    \begin{subfigure}[t]{0.23\linewidth}
        \caption{\label{fig: STminus -60 mV}}
        \includegraphics[width=\linewidth]{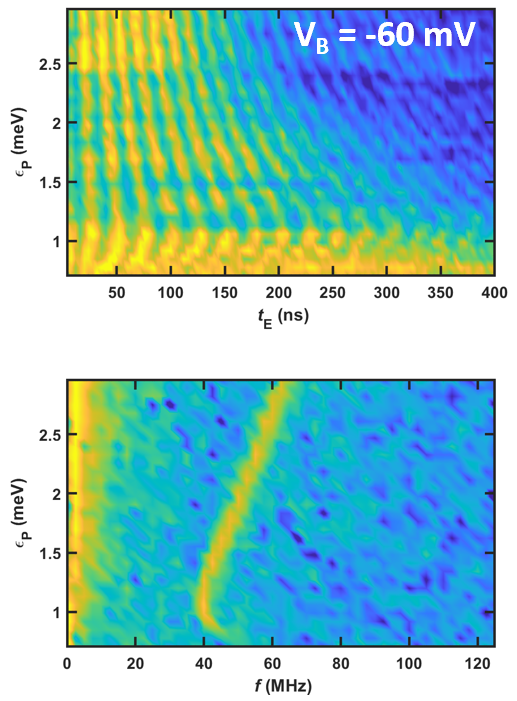} 
    \end{subfigure}
    \begin{subfigure}[t]{0.23\linewidth}
        \caption{\label{fig: STminus -55 mV}}
        \includegraphics[width=\linewidth]{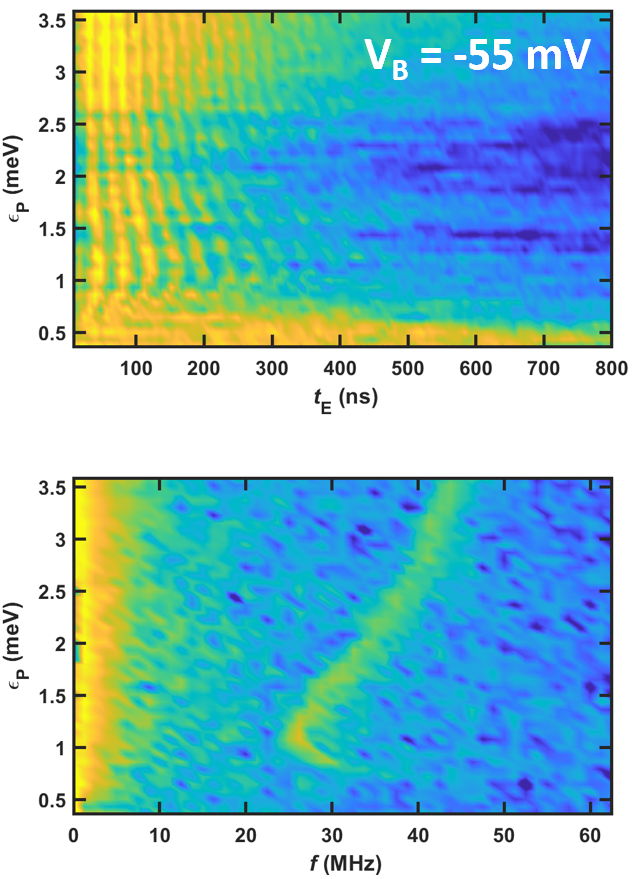} 
    \end{subfigure}
    \begin{subfigure}[t]{0.23\linewidth}
        \caption{\label{fig: STminus -54 mV}}
        \includegraphics[width=\linewidth]{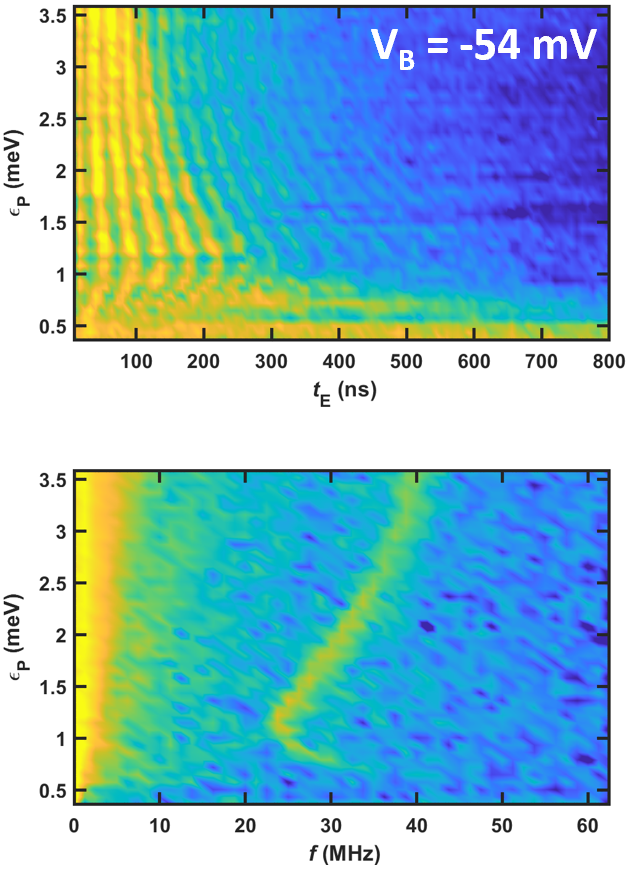} 
    \end{subfigure}
    \begin{subfigure}[t]{0.05\linewidth}
        \raisebox{-3.6cm}{%
        \includegraphics[scale=0.6]{STminusOscillationsVb_colorbar}}
    \end{subfigure}

    \begin{subfigure}[t]{0.23\linewidth}
        \caption{\label{fig: STminus -53 mV}}
        \includegraphics[width=\linewidth]{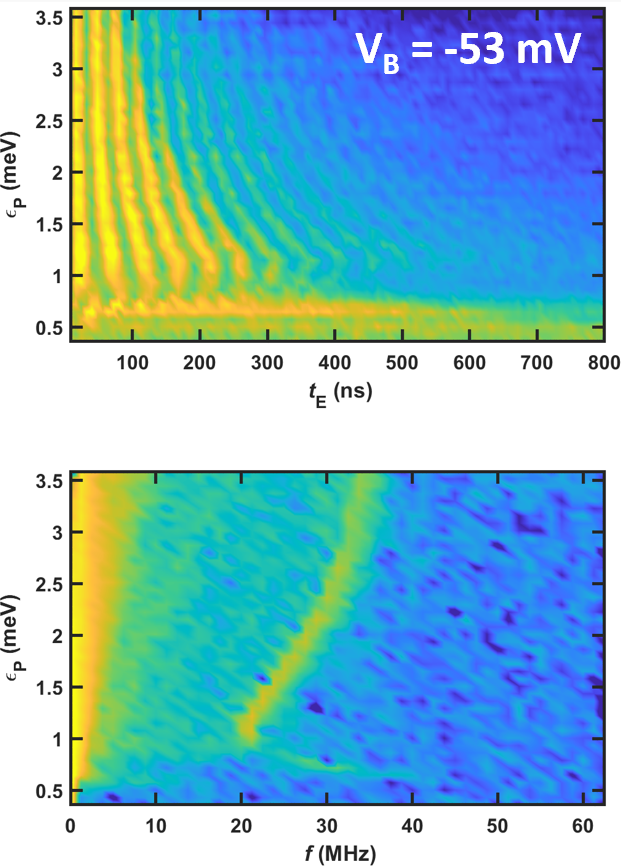} 
    \end{subfigure}
    \begin{subfigure}[t]{0.23\linewidth}
        \caption{\label{fig: STminus -52 mV}}
        \includegraphics[width=\linewidth]{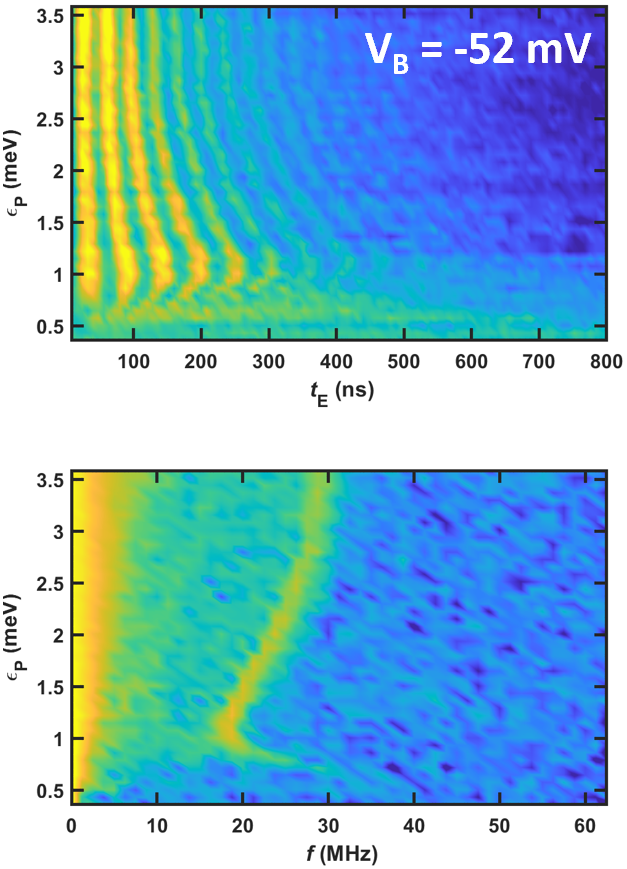}
    \end{subfigure}
    \begin{subfigure}[t]{0.23\linewidth}
        \caption{\label{fig: STminus -51 mV}}
        \includegraphics[width=\linewidth]{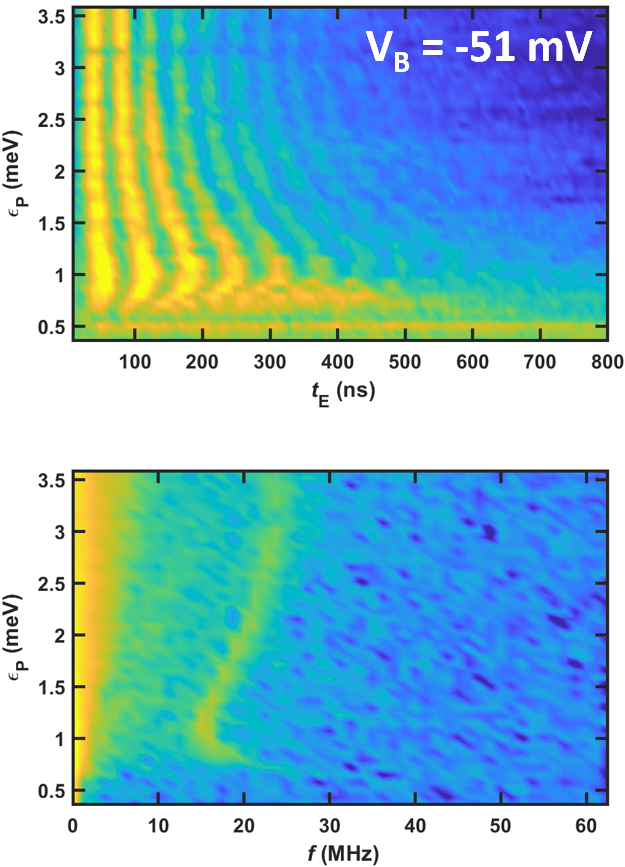} 
    \end{subfigure}
    \hspace{0.05\linewidth}

    \begin{subfigure}[t]{0.23\linewidth}
        \caption{\label{fig: STminus -50 mV}}
        \includegraphics[width=\linewidth]{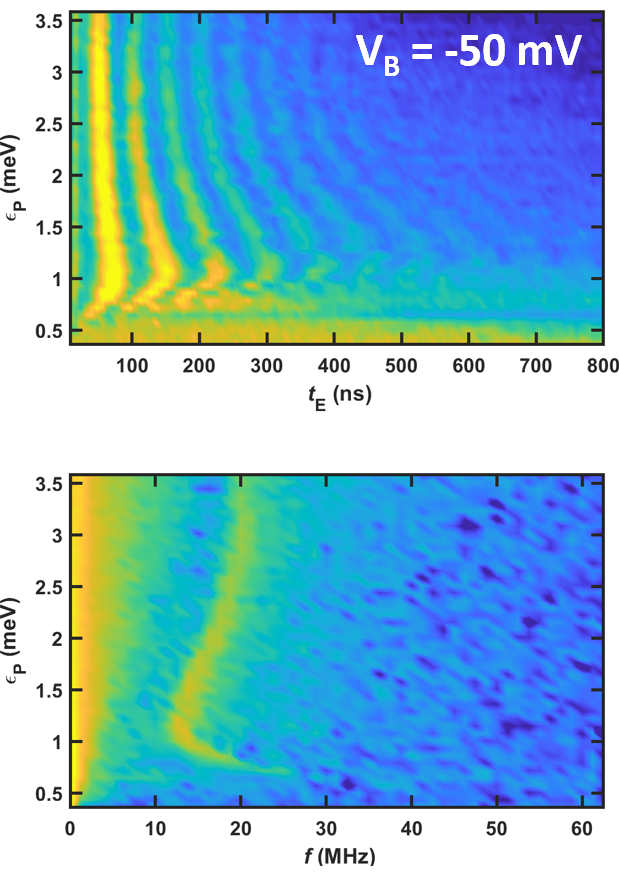} 
    \end{subfigure}
    \begin{subfigure}[t]{0.23\linewidth}
        \caption{\label{fig: STminus -49 mV}}
        \includegraphics[width=\linewidth]{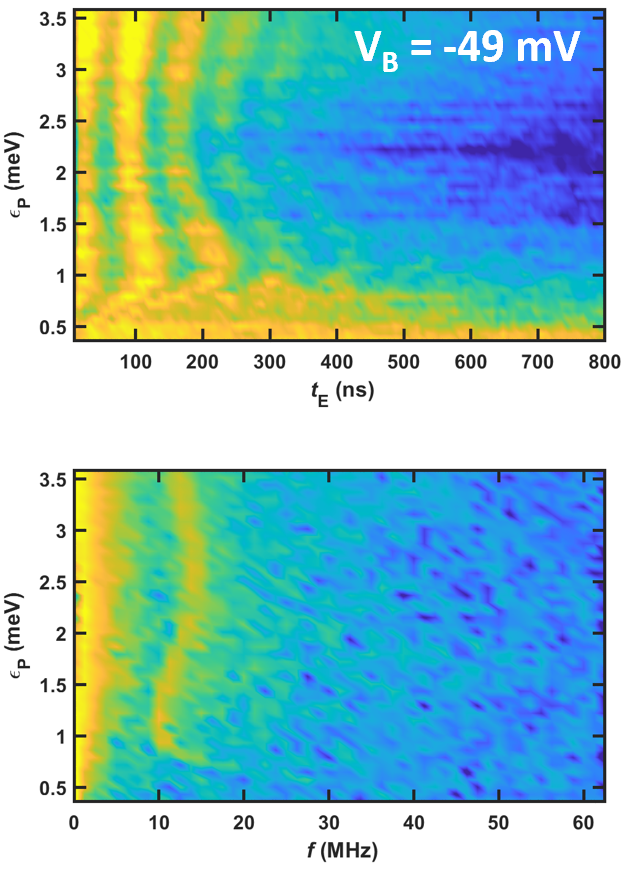} 
    \end{subfigure}
        \begin{subfigure}[t]{0.23\linewidth}
        \caption{\label{fig: STminus -48 mV}}
        \includegraphics[width=\linewidth]{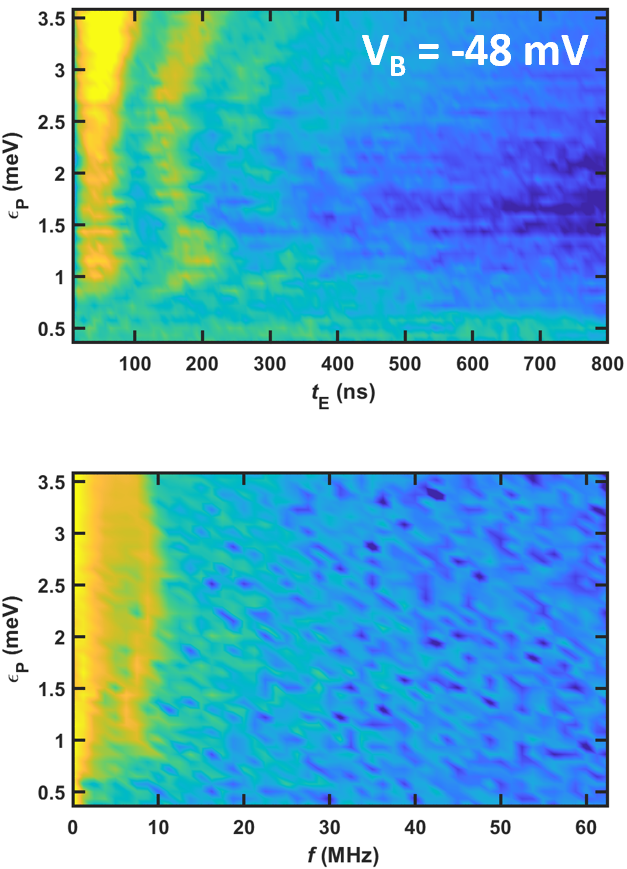} 
    \end{subfigure}
    \hspace{0.05\linewidth}
\caption{\label{fig: STminus vs B}{\textbf{(a)-(i)} Evolution of the $S-T\_$ oscillations (upper panel) and their corresponding FFTs (lower panel) as a function of the middle barrier gate voltage $V_B$. Applying a more positive barrier gate voltage decreases the frequency of the oscillations throughout the entire detuning range. Both the minimum and maximum frequencies decrease as $V_B$ becomes more positive, indicating a reduction in both $\Delta_{ST\_}$ and $\overline{g}$.}}
\end{figure*}

We now focus on modulating the coherent evolution of the $S-T\_$ states by adjusting the voltage applied to the barrier separating the two quantum dots. Over a small range of voltage (12 mV), Fig. \ref{fig: STminus vs B} illustrates the dramatic transformation the $S-T\_$ oscillations undergo. As $V_B$ increases, it is clear the frequency of evolution between the two states decreases monotonically. From the Fourier transform of these oscillations, we can isolate two quantities of interest, namely $\overline{g}$ from the frequency at large detuning following $f \sim \overline{E}_z/h$ and $\Delta_{ST\_}$ from the minimum frequency near $\epsilon_P=1$ meV, where $f = 2\Delta_{ST\_}/h$.

We would like to note that the location of the frequency minimum $\epsilon_*$ is determined by the tunnel coupling $t_c$ and $\overline{E}_z$ from the condition $J=\overline{E}_z$ \cite{Jirovec2022}:

\begin{equation}
\label{eq:epsilon*}
\epsilon_* =  \frac{2t_c^2-\overline{E}_z^2}{\overline{E}_z},
\end{equation}

\noindent Because $\epsilon_*$ remains approximately constant throughout this range of $V_B$, a decrease in $t_c$ must be accompanied by a decrease in $\overline{E}_z$. While it is evident from the sharper rises seen in the FFTs of Fig. \ref{fig: STminus vs B} that $t_c$ decreases with $V_B$, a similar decline in $\overline{E}_z$, and therefore $\overline{g}$, is necessarily present. From the $S-T\_$ evolution frequency at large $\epsilon_P$ in Fig. \ref{fig: STminus vs B}, we can then extract the dependence of $\overline{g}$ on $V_B$. 

\section{\label{sec: discussion} Discussion}

\begin{figure}[h]
    \begin{subfigure}[b]{0.45\columnwidth}
        \caption{\label{fig: dSTminus vs barrier}}
        \includegraphics[width=\linewidth]{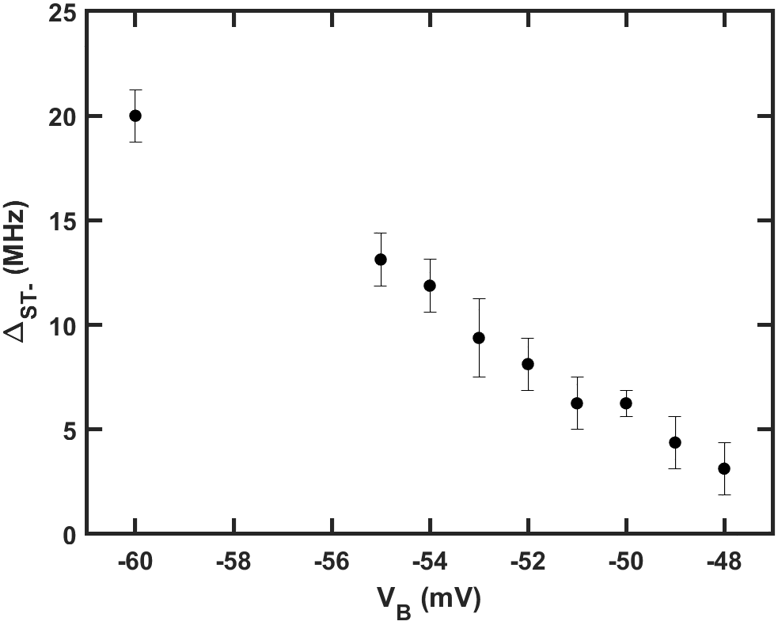}  
    \end{subfigure}
    \begin{subfigure}[b]{0.45\columnwidth}
        \caption{\label{fig: gbar vs barrier}}
        \includegraphics[width=\linewidth]{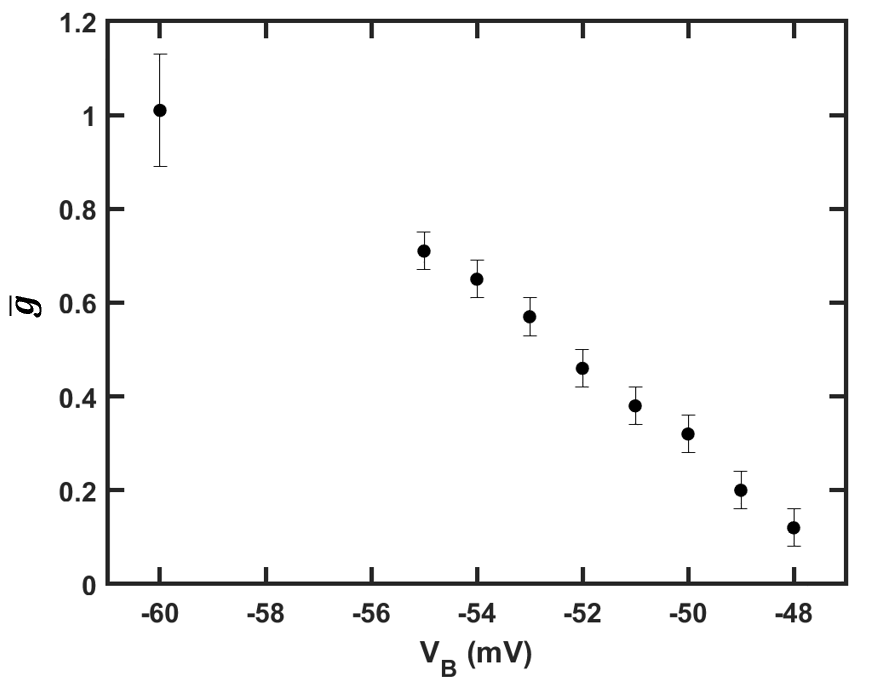}
    \end{subfigure}

    \begin{subfigure}[b]{0.9\columnwidth}
        \caption{\label{fig: strain cartoon}}
        \includegraphics[width=\linewidth]{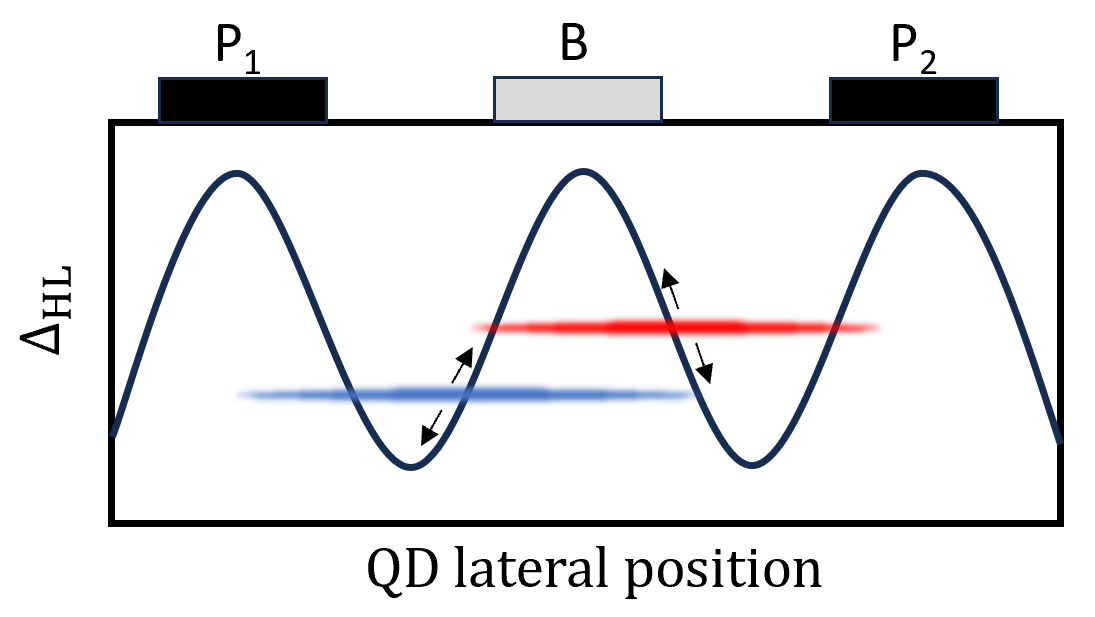}
    \end{subfigure}
\caption{\label{fig: dST and gbar vs barrier}{\textbf{(a)} $\Delta_{ST\_}$ as a function of barrier voltage. Values are extracted from the minimum frequency of the FFTs shown in Fig. \ref{fig: STminus vs B}. \textbf{(b)} $\overline{g}$ is extracted from frequency at large detunings in Fig. \ref{fig: STminus vs B} and plotted versus the barrier voltage. Both parameters show a strong linear dependence on the barrier gate voltage. This behavior can be explained by the dots moving through a non-uniform strain environment, which directly impacts the $g$-factors of each dot. Error bars for (a) and (b) are calculated from the linewidth of the Fourier transform data. \textbf{(c)} Cartoon depicting the $\Delta_{HL}$ profile underneath the confinement gates due to the effects of strain.}}
\end{figure}

Fig. \ref{fig: dST and gbar vs barrier} demonstrates the linear trend both $\Delta_{ST\_}$ and $\overline{g}$ follow with respect to the barrier voltage $V_B$. Recall the $S-T\_$ coupling is determined by both the spin-orbit coupling and the effect of the anisotropic $g$-tensors, $g_a$. We assert it is the latter of these contributions that is affected over the small range of $V_B$ considered here. Consequently, to justify the modulation of the qubit frequency in Fig. \ref{fig: dST and gbar vs barrier}, we seek a mechanism that simultaneously reduces both $\overline{g}$ and $g_a$ as $V_B$ increases. We will argue these changes are accomplished through increasing the admixture of light-hole (LH) states into the predominantly heavy-hole (HH) ground state of the Ge quantum dot. 

While it is well known the upper valence bands in Ge/SiGe heterostructures are composed primarily of HH states due to a large HH-LH splitting $\Delta_{HL}$ \cite{Scappucci2021}, an accurate understanding of the $g$-tensor in many Ge materials requires the consideration of the LH bands \cite{Drichko2014, Watzinger2016, Ares2013}. To understand the consequences of this mixing, it is beneficial to first examine the $g$-factor components of both bands in the case of bulk Ge. As described in Refs. \cite{Kiselev2001,Nenashev2003,Watzinger2016}, for the pure HH state, the out-of-plane $g$-factor is $g_\perp = 6 \kappa + \frac{27q}{2}$ and the in-plane component is $g_\parallel = 3q$, where $\kappa = 3.41$ and $q=0.07$ are the magnetic Luttinger parameters. Note this $\kappa$ and $q$ result in the large anisotropy of the $g$-tensor: $g_\perp \gg g_\parallel$. Conversely, for pure LH states, $g_\perp = 2\kappa$ and $g_\parallel = 4\kappa$. Comparing these two bands, the LH state has a smaller $g_\perp$ but greater $g_\parallel$ compared to the HH state. Therefore, when increasing the LH admixture in the ground state of the quantum dot, we expect a decrease in $g_\perp$ and an increase in $g_\parallel$, which has been experimentally observed for various mixing mechanisms \cite{Nenashev2003,Ares2013,Watzinger2016,Jirovec2022}. 

Due to the large anisotropy between $g_\perp$ and $g_\parallel$, $\overline{g}$ is dominated by its out-of-plane component $\overline{g}_\perp$. With an increase in HH-LH mixing, we then expect a decrease in $\overline{g}$ through a reduction in $g_\perp$ for either dot. On the other hand, the in-plane $g$-factors are the leading order terms defining $g_a$, where a more similar $g_\parallel$ between the two dots diminishes $g_a$. Increasing the HH-LH mixing decreases $g_a$ when the in-plane $g$-factors change in such a way that the difference between $g_\parallel$ for the left and right dot lessens. Importantly, this mixing mechanism can then explain both downward trends we observe in Fig. \ref{fig: dST and gbar vs barrier}.
 
Although a complete theoretical description lies outside the scope of this work, we will now discuss how a non-uniform strain profile is a viable candidate for this rise in HH-LH mixing as $V_B$ is varied. Strain originates from the differences in thermal contraction between the gate electrodes defining the quantum dots and the substrate. This strain can both alter $\Delta_{HL}$ and directly mix the HH and LH states, where these effects are greatest along the edges of the confinement gates \cite{Nakaoka2004,Liles2021,Corley-Wiciak2023}. In one case, Corley-Wiciak et al., 2023 measured the strain profile of a gate-defined quantum dot device and simulated that $\Delta_{HL}$ can vary as much as 4$\%$ \cite{Corley-Wiciak2023}. Although this percentage seems small at first glance, the degree of mixing between the HH and LH states scales as $\left(\frac{1}{\Delta_{HL}}\right)^2$ \cite{Luo2015}, where even admixtures of 1$\%$ significantly reduce $g_\perp$ \cite{Watzinger2016}.  

From a rough calculation of the quantum dot positions as a function of $V_B$ (see section S\ref{sec: dot position}) and using the results of Ref. \cite{Corley-Wiciak2023} for a qualitative picture, we estimate an overall shift of the right dot's position away from the middle barrier leads to a $\sim 3\%$ decrease in $\Delta_{HL}$ as the dot moves into a region of increased strain (see Fig. \ref{fig: strain cartoon}). With this change leading to a $9\%$ enhancement in the LH admixture, we can expect a decrease in both $\bar{g}$ and $\Delta_{ST\_}$. We want to stress these values are crude calculations and only serve as a guide to how the hole $g$-tensor evolves with respect to $V_B$ and explain the trends in Fig. \ref{fig: dST and gbar vs barrier}. 

\section{\label{sec: conclusion} Conclusion}

In summary, we have explored the coherent oscillations in a Ge hole double dot between the singlet, $|S\rangle$, and polarized triplet state, $|T\_\rangle$. The dephasing time of this manipulation strongly depends on the operation detuning with a maximum of $T_2^* = 600$ ns, while the spin relaxation time at the readout point was measured to be $T_1 = 17.2$ $\mu$s. The maximum in $T_2^*$ coincides with the minimum in the $S-T\_$ energy splitting, where the system is insensitive to the noise disturbing $J$ and $\overline{E}_z$. Furthermore, we observe the frequency of evolution between these spin states can be modulated through the voltage of the middle barrier separating the two dots. The frequency dependence on $V_B$ points to the changing dot position over a variable strain profile as the reason for adjusting the qubit frequency. These results suggest strain can be exploited to fine-tune qubit frequencies in Ge. Furthermore, if a variable frequency profile is not desired, the sensitivity of the $g$-tensor to the quantum dot position can be mitigated by reducing strain in the system, such as by defining gate electrodes with palladium instead of gold to closer match the thermal response of Ge \cite{Mooy2022}. 

\section*{\label{sec: acknowledgements} Acknowledgements}

Research was sponsored by the Army Research Office (ARO) and was accomplished under Grant No. W911NF-23-1-0016 (at UCLA) and Grant No. W911NF-22-S-0006 (at TU Delft). The views and conclusions contained in this document are those of the authors and should not be interpreted as representing the official policies, either expressed or implied, of the Army Research Office (ARO), or the U.S. Government. The U.S. Government is authorized to reproduce and distribute reprints for Government purposes notwithstanding any copyright notation herein. G.S. also acknowledges support through two “Projectruimte”, associated with the Netherlands Organization of Scientific Research (NWO).

\section*{\label{sec: device fabrication} Device Fabrication}

The device was fabricated on top of a Ge/SiGe heterostructure with the strained Ge quantum well buried 55 nm beneath the surface. Ohmic regions were patterned with photolithography and the wafer was dipped in buffered HF (BOE) to remove the thin capping layer of SiO$_2$. These regions were then metallized with 60 nm of Pt through e-beam evaporation. We used e-beam lithography to pattern all device leads and a second e-beam evaporation to deposit 5/45 nm of Ti/Au. A 100 nm insulating layer of Al$_2$O$_3$ was grown through atomic layer deposition, and a subsequent global top gate was patterned with photolithography. This was followed by a final e-beam evaporation of 100 nm of Al to form the global accumulation gate. The device then underwent a forming gas anneal at 420 C for 1 hour to repair defects in the oxide and anneal the Pt Ohmic regions into the substrate. 

\section*{\label{sec: measurement setup} Measurement setup}

The device was cooled in a Triton dilution refrigerator with a base temperature of 48 mK. All SET current measurements were performed with an SR 830 Lock-in amplifier. When measuring stability diagrams, the lock-in excitation voltage was applied to both plungers and the current through the SET was fed back into the lock-in for integration and demodulation. A voltage pulse to each plunger was supplied by a Tektronix AWG 610 with its pulse frequency modulated by the lock-in. With the lock-in integration time set to 100 ms, ~5000 pulse sequences were averaged for each data point during the spin manipulation measurements. All measurements using the AWG pulse were made with no lock-in excitation voltage applied to the plungers. The magnet used was the model B444-N52 produced by K\&J Magnetics, Inc. and attached directly next to the device's PCB. A Hall probe measured the magnetic field at the device's position on the PCB.          

\bibliography{references} 

\newpage
\widetext
\begin{center}
\textbf{\large Supplementary Materials}
\end{center}

\setcounter{section}{0}
\setcounter{equation}{0}
\setcounter{figure}{0}
\setcounter{table}{0}
\setcounter{page}{1}
\makeatletter
\renewcommand{\theequation}{S\arabic{equation}}
\renewcommand{\thefigure}{S\arabic{figure}}
\renewcommand{\bibnumfmt}[1]{[#1]} 
\renewcommand{\citenumfont}[1]{#1} 

\section{\label{sec: AWG sequence} AWG Pulse Sequence}

\begin{figure}[h]
    \begin{subfigure}[t]{0.45\columnwidth}
        \caption{\label{fig: pulse sequence}}
        \includegraphics[width=\linewidth]{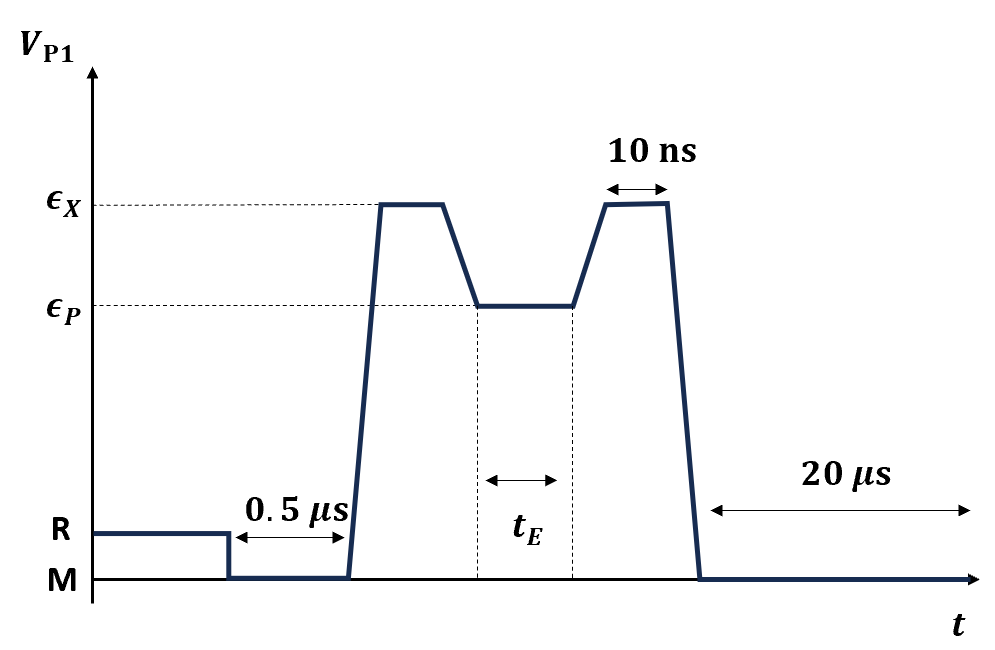}
    \end{subfigure}
    \begin{subfigure}[t]{0.35\columnwidth}
        \caption{\label{fig: pulse sequence stability diagram}}
        \includegraphics[width=\linewidth]{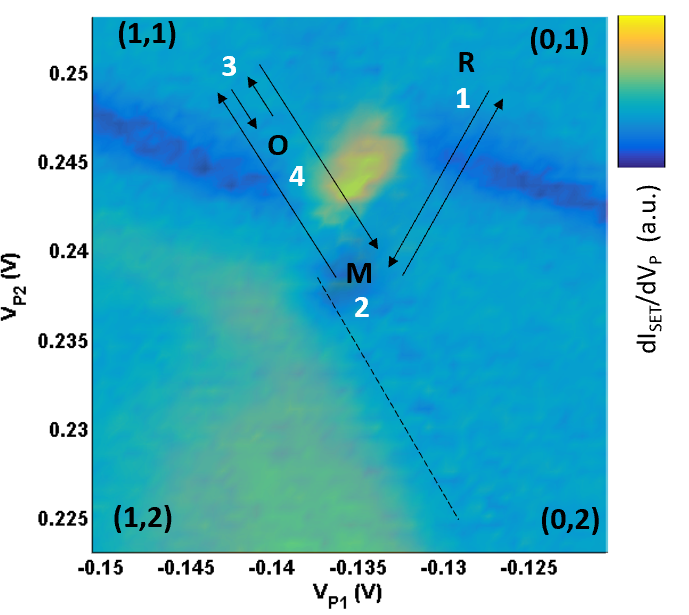}
    \end{subfigure}
\caption{\label{fig: AWG pulse sequence}{\textbf{(a)} AWG pulse applied to P$_1$ to observe coherent evolution between the singlet and triplet states (not to scale). The same pulse but inverted (excluding step R) is applied to P$_2$. All ramp times were set to 1 ns. \textbf{(b)} Two stability diagrams overlaid on top of each other--one scan made with the AWG pulse on and the other scan without the AWG pulse. A dark blue spot appears below the interdot transition line with the pulse on, which we use to mark the measurement position M. Numbers denote the sequence the system undergoes with the AWG pulse. (1) The state is first reset at R for 0.5 $\mu$s, then (2) moved to M for initialization (also 0.5 $\mu$s). (3) The singlet is pulsed deep into (1,1) and allowed to evolve for 10 ns. (4) This ramp is followed by evolution at point O where the system idles at $\epsilon_P$ for time $t_E$. A final rotation is performed at (3) to project the state back onto the singlet and triplet states, where readout is performed at (M) for 20 $\mu$s. Finally, the system is reset at R before undergoing this cycle again.}}
\end{figure}

Fig. \ref{fig: AWG pulse sequence} outlines the AWG voltage pulse applied to the plungers to manipulate the hole configuration between points R (reset), M (measurement and initialization), and O (operation). To coherently manipulate the two-hole state, we first reset the system at R for 1 $\mu$s, allowing holes to tunnel off both dots until the (0,1) configuration is reached. A hole was then loaded into the right dot by waiting at M for 0.5 $\mu$s to initialize the system into the singlet ground state S(0,2). The $|S\rangle$ state was generated by quickly pulsing to $\epsilon_X$ with a 1 ns rise time. From this ramp and a 10 ns idle period at $\epsilon_X$, a mixture between $|S\rangle$ and $|T\_\rangle$ is created. The system is then moved to the operation point at $\epsilon_P$ for an evolution time $t_E$ before the process was reversed to arrive back at M for measurement of the final state.  

To read out the final two-hole state, the system was held at point M for 20 $\mu$s. Point M sits inside the Pauli spin blockade (PSB) region, enabling spin to charge conversion when reading out the spin state of the double hole system. Due to spin conservation while in the PSB region, the SET current will either detect a change in the hole configuration to (0,2), signifying the final state as $|S\rangle$, or no change in hole occupation will be seen, indicating the final state as $|T\_\rangle$. 

\newpage
\section{\label{sec: T2*} $T_2^*$ dephasing}

\begin{figure}[h]
\includegraphics[width=\linewidth]{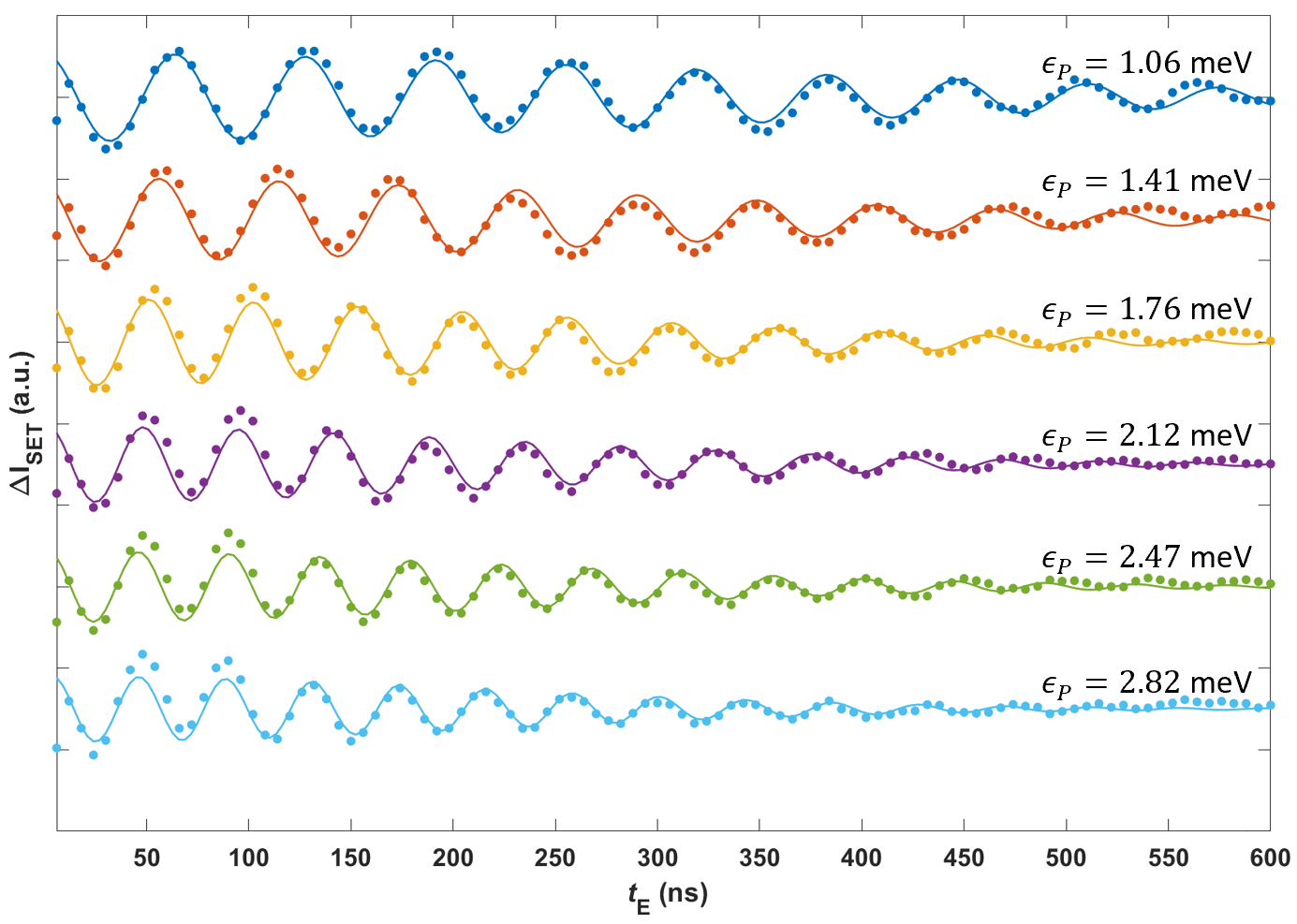}
\caption{\label{fig: T2* linecuts}{Traces taken from Fig. \ref{fig: STminus -51 mV} for several values of operation detuning $\epsilon_P$. Each trace is fit to a decaying sinusoid with an exponential background and offset. The background is removed before plotting the change in the SET current ($\Delta I_{SET}$) for both the trace and fit (solid line). Each trace is offset along the vertical axis for clarity.}}
\end{figure}

To measure $T_2^*$, we extracted the decay of oscillations between $|S\rangle$ and $|T\_\rangle$ in Fig. \ref{fig: STminus oscillations}. Each linecut along $\epsilon_P$ is fit to the equation $P=Ae^{-(t/T_2^*)^2}\cos{(\omega t+B)}+Ce^{-t/D}+E$, where $A, B, C, D,$ and $E$ are fitting parameters in addition to $T_2^*$. The angular frequency $\omega=2\pi f$ is already known from the Fourier transform data. Several fitted traces are shown in Fig. \ref{fig: T2* linecuts} for Fig. \ref{fig: STminus -51 mV}, where the background $(Ce^{-t/D} + E)$ has been subtracted off. A clear dependence of the dephasing time on detuning can be seen, where $T_2^*$ continues to decrease past $\epsilon=1$ meV.   
 
As discussed in the main text, the dependence of $T_2^*$ on $\epsilon_P$ can be explained by contributions from electrical and magnetic noise affecting the Zeeman terms in our Hamiltonian (Eqn. \ref{eq: H}). For detunings on either side of the $S-T\_$ anticrossing at $\epsilon_*$, $J - \overline{E}_z$ sets the $f$ of evolution. In this region, the noise affecting $t_c$ and $\overline{E}_z$ dominates. We want to note for $\epsilon \ll \epsilon_*$, charge noise disturbing the exchange term $J$ governs $T_2^*$; however, that regime was not measured here. For $\epsilon \gg \epsilon_*$, the $S-T\_$ energy levels become increasingly parallel to each other. Once again, $\overline{E}_z$ determines the energy splitting and consequently $\delta \overline{E}_{z_\text{rms}}$ is the leading noise source.     

Near $\epsilon_*$, the evolution frequency reaches a minimum as the exchange and average Zeeman energy balance each other out, leaving $f=\frac{2 \Delta}{h}$ and $\delta \Delta_\text{rms}$ setting $T_2^*$. Because $\delta \Delta_\text{rms} < \delta \overline{E}_{z_\text{rms}}$, a peak is seen in $T_2^*$ at $\epsilon^* =  1$ meV in Fig. \ref{fig: T2* plot}.    
 
\section{\label{lever arms} Lever arm measurement}

\begin{figure}[h]
\includegraphics[width=0.45\linewidth]{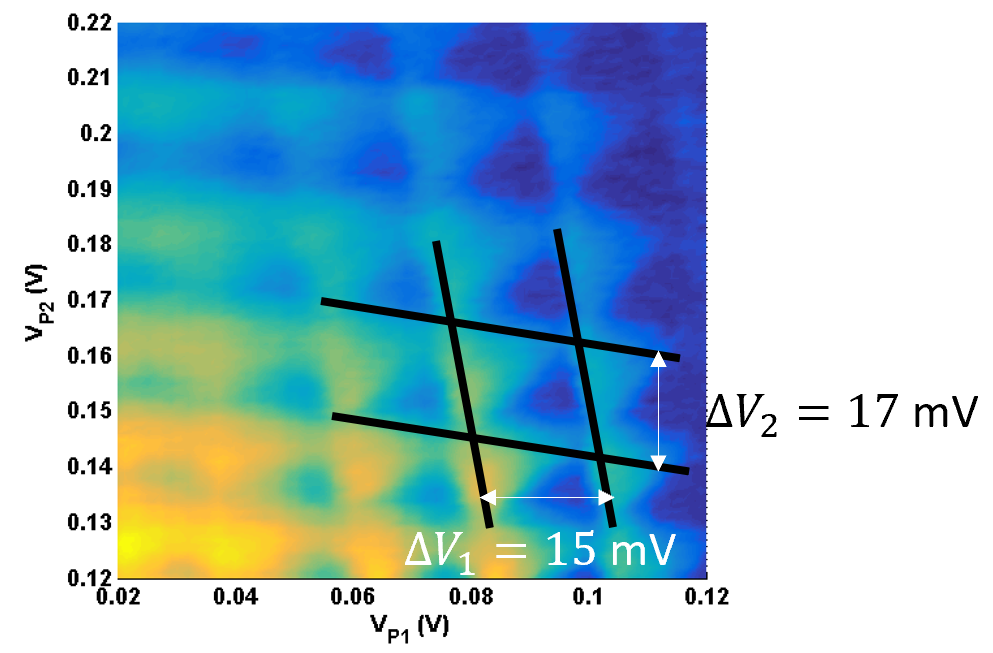}
\caption{\label{fig: lever arms}{Stability diagram illustrating transport through both quantum dots with a bias voltage $V_{SD} = 1$ mV across the source-drain reservoirs. Triangles can be seen where the dot charging lines intersect. The dimensions of these triangles are directly related to the source-drain bias and plunger voltage, which we used to calculate the lever arm of each plunger: $\alpha_1 = 0.12$ and $\alpha_2 = 0.11$ eV/V.}}
\end{figure}

When a small bias voltage $V_{SD}$ is applied across the reservoirs of the double quantum dot, conductance regions in the shape of triangles will form at the intersections of charge transitions between the two dots \cite{vanderwiel2003}. The dimensions of these triangles are directly related to the energy separation between the source and drain, $eV_{SD}$, and the chemical potential of each dot being controlled by its respective plunger voltage:
\begin{equation*}
    \alpha \Delta V_g = |eV_{SD}|,
\end{equation*}

\noindent where $\alpha$ is the lever arm relating the plunger voltage to the dot chemical potential. By measuring the lengths of these triangles along each plunger voltage axis (Fig. \ref{fig: lever arms}), we determined the conversion factor $\alpha$ from the ratio of plunger voltage to energy: $\alpha=eV_{SD}/\Delta V_P$. We find the lever arms for P$_1$ and P$_2$ to be $\alpha_1 = 0.12$ and $\alpha_2 = 0.11$ eV/V, which were used to convert the plunger voltages to energies for all calculations in this work.

\newpage
\section{\label{fast oscillations} Fast $S-T\_$ oscillations}

\begin{figure}[h]
    \begin{subfigure}[t]{0.45\columnwidth}
        \caption{\label{fig: fast oscillations scan}}
        \includegraphics[width=\linewidth]{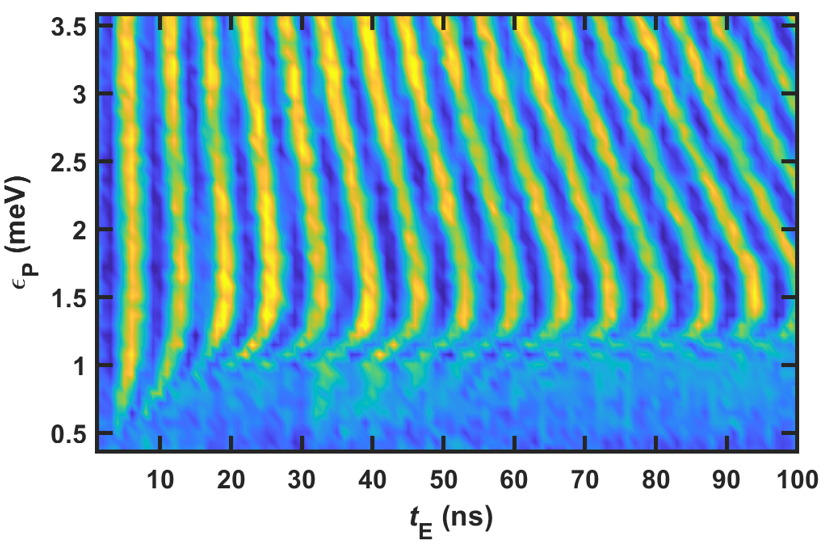}
    \end{subfigure}
    \begin{subfigure}[t]{0.45\columnwidth}
        \caption{\label{fig: fast oscillations FFT}}
        \includegraphics[width=\linewidth]{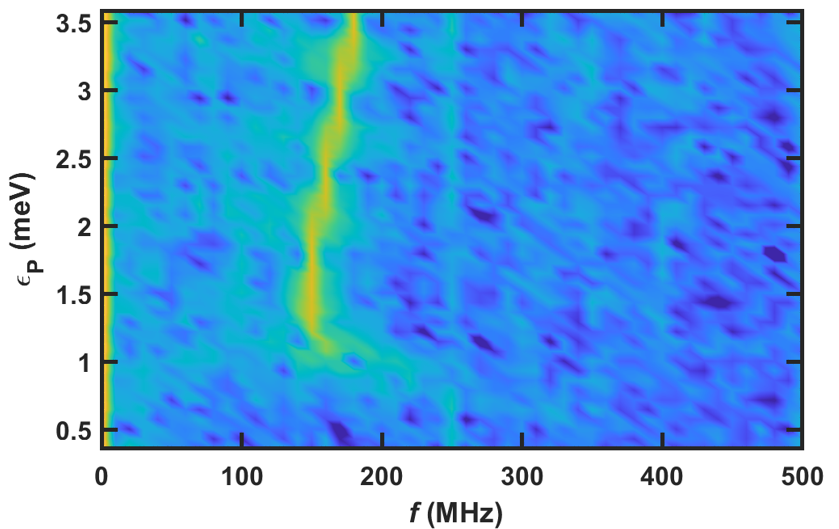}
    \end{subfigure}
\caption{\label{fig: fast oscillations}{\textbf{(a)} A scan of the singlet-triplet evolution illustrating the highest frequency oscillations seen from our data set. \textbf{(b)} The the FFT of (a). The $S-T\_$ anticrossing gap is $\Delta_{ST\_}=75$ MHz, while the frequency at large detunings reaches 180 MHz, signifying $\overline{g} = 2.8$.}}
\end{figure}

The fastest oscillations we observed are shown in Fig. \ref{fig: fast oscillations}. This scan was taken at a center barrier voltage of $V_B=-180$ mV, and illustrates $\Delta_{ST\_}$ reaching 75 MHz and the maximum $S-T\_$ energy splitting surpassing 180 MHz, which corresponds to a maximum $\overline{g} \approx 2.8$. While this $\overline{g}$ is larger than those reported in the main text, we do not expect $\overline{g}$ to grow indefinitely as we decrease the middle barrier voltage. As $V_B$ draws the two dots closer to the middle barrier's edge, the continuous increase in tunnel coupling $t_c$ between the two dots will eventually make system inoperable as a singlet-triplet qubit.

\newpage
\section{\label{sec: ST(0,2) splitting} Singlet-Triplet splitting in (0,2) for readout}

The energy splitting between the singlet and triplet states in the (0,2) charge configuration was measured by varying the readout position $\epsilon_r$ of the Ramsey pulse. As the readout position is further pushed deeper into the (0,2) charge configuration, it will eventually surpass the Pauli spin blockade (PSB) boundary the SET uses to distinguish the singlet and triplet states. While in the PSB region, only the singlet (1,1) state is energetically allowed to tunnel to the singlet (0,2) state. In contrast, all triplet states are prohibited from tunneling to (0,2) due to large singlet-triplet splitting $E_{ST_{02}}$ in the (0,2) charge configuration. The triplet states also are prevented from tunnelling to the singlet (0,2) state due to the conservation of spin. 

\begin{figure}[h]
    \begin{subfigure}[t]{\linewidth}
        \includegraphics[width=\linewidth]{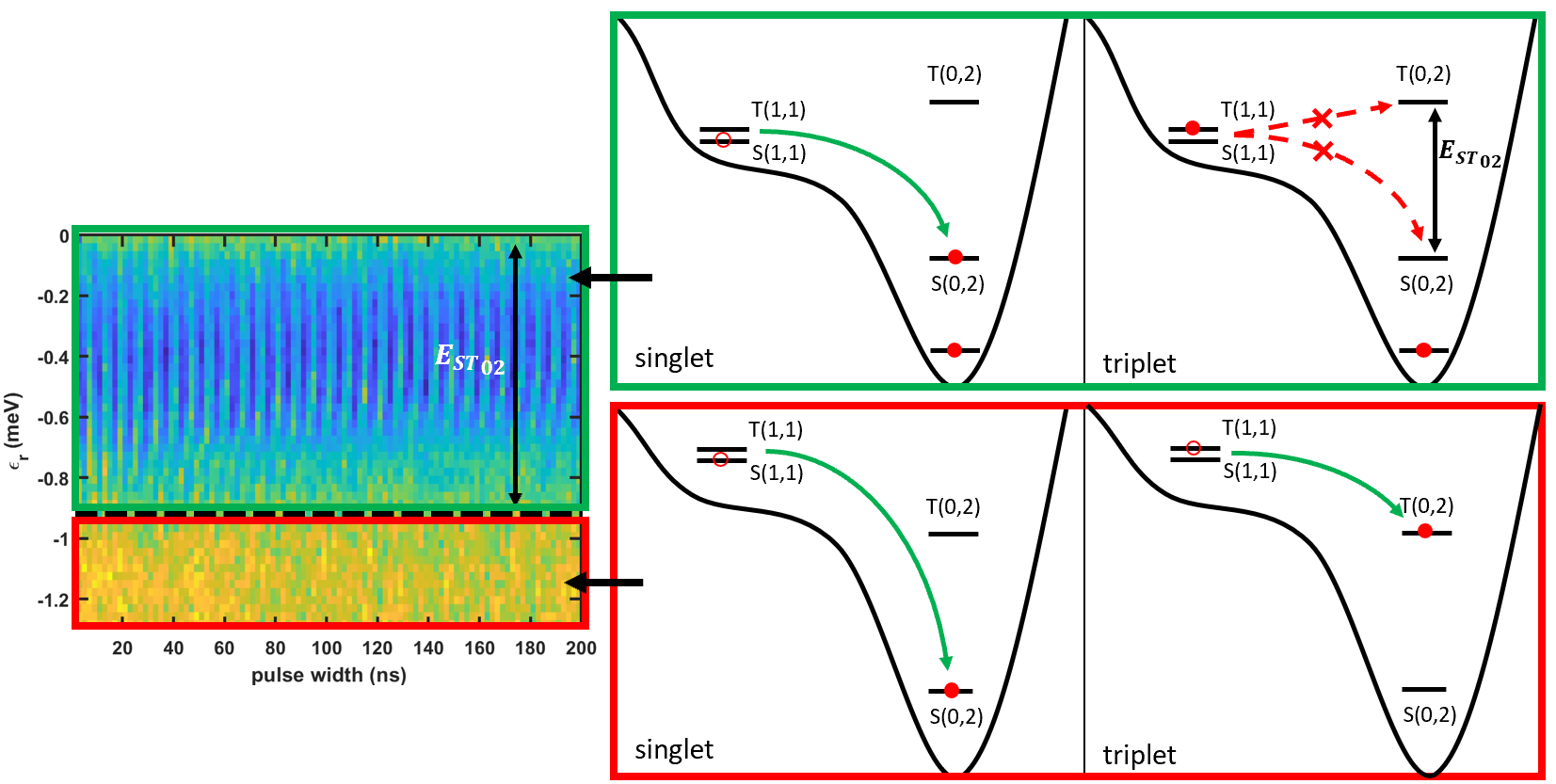}
    \end{subfigure}
\caption{\label{fig: ST splitting}{By scanning the readout position $\epsilon_r$ of the Ramsey pulse, we measured the energy splitting between the singlet and triplet states in the (0,2) charge configuration. As the readout position is moved further into (0,2), PSB is eventually lifted as the (1,1) triplet state is allowed to tunnel to the (0,2) triplet. By detecting the readout position where this transition occurs, we measured $E_{ST_{02}} = 0.9 $ meV.}}
\end{figure}

However, When the readout position is sufficiently pushed deep enough into the (0,2) charge configuration, that is $\epsilon_r>E_{ST_{02}}$, it will be energetically allowed for both the singlet and triplet (1,1) states to tunnel to their respective (0,2) configurations while conserving spin. At this point, the readout position is outside of the PSB region and the SET will no longer distinguish the singlet from the triplet states. From Fig. \ref{fig: ST splitting}, we measure this cutoff and therefore $E_{ST_{02}}$ to be 0.9 meV.  

Knowing the orbital energy difference defines the (0,2) singlet-triplet splitting, we can use $E_{ST_{02}}$ to estimate the size of the right dot by approximating the confinement potential to be a 2d box and the dot shape to be a disk: 

\begin{align*}
    E_{ST_{02}} &= \frac{3\hbar^2 \pi^2}{2m^* L^2}\\
    E_{ST_{02}} &= \frac{3\hbar^2 \pi^2}{2m^* \pi r^2},
\end{align*}

\noindent where $m^* = 0.09 m_e$ and $r$ is the radius of the dot. From $E_{ST_{02}} = 0.9$ meV, we calculate the radius of the right dot to be $r=66$ nm. By comparing the charging energies of the left and right dot and using the fact that $E \sim 1/r^2$, we further estimate the size of the left dot to be $r=75$ nm. 

\newpage
\section{\label{sec: dot position} Estimation of $\Delta_{HL}$ from dot positions}

\begin{figure}[h]
    \begin{subfigure}[t]{\linewidth}
        \includegraphics[width=0.6\linewidth]{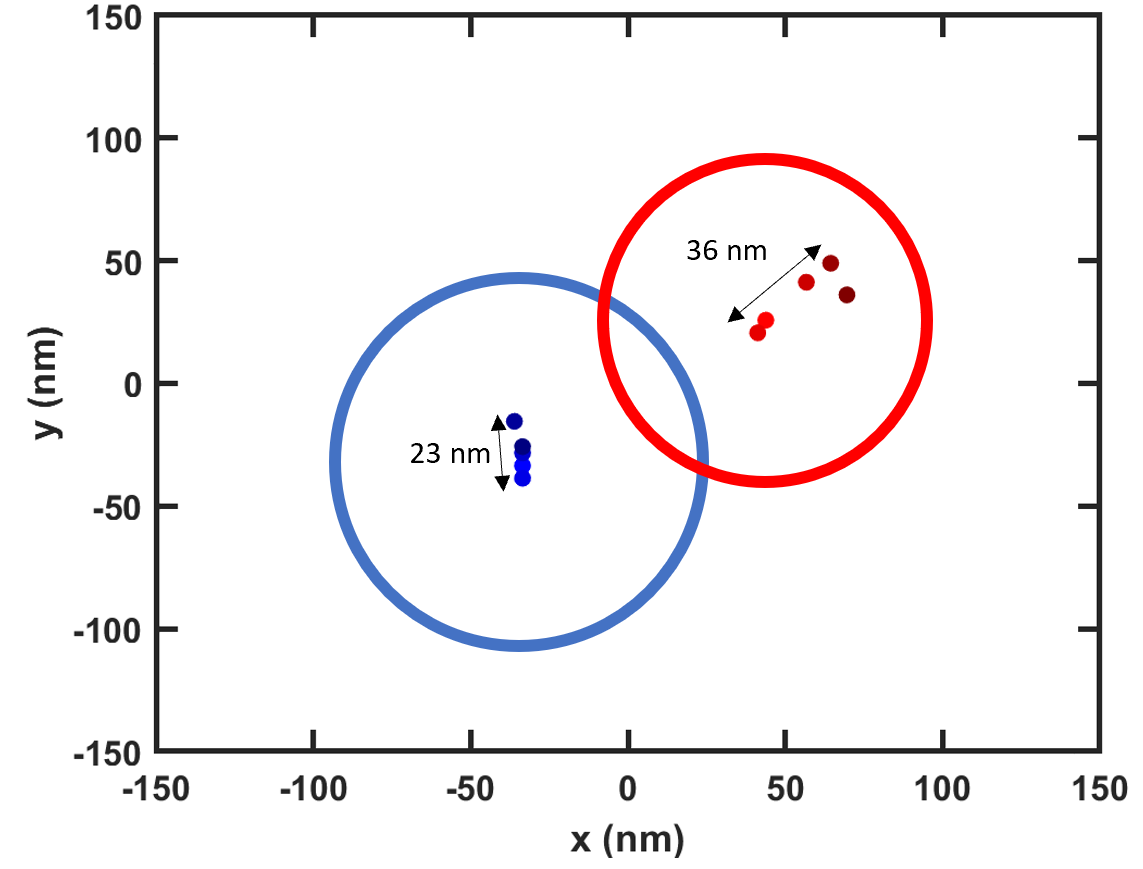}
    \end{subfigure}
\caption{\label{fig: dot positions}{The positions of both quantum dots are estimated from their capacitance ratios to the two plungers and center barrier. The coordinate $x=0$ is where the center barrier is placed. Blue (red) dots correspond to the left (right) dot positions. The darker shades correspond to more positive $V_B$. It is evident that as $V_B$ increases, the right dot tends to shift away from the center barrier, while the left dot fluctuates around its starting position. The diameters of the two quantum dots are depicted as circles and estimated in section S\ref{sec: ST(0,2) splitting}.}}
\end{figure}

We were able to make a crude estimate of the quantum dot positions using capacitance ratios of the quantum dot to the two plungers and barrier gates. The capacitance ratios were found from the slope of the charging lines in the stability diagram. From the relative strength of the effect two gates have on a dot and knowing the gate positions in the x-y plane, we can create what is known as an Apollonian circle. Creating Apollonian circles for two different pairs of gates allowed us to estimate the dot position where these two circles intersect. Repeating this process for various $V_B$ generated the data in Fig. \ref{fig: dot positions}. 

From Fig. \ref{fig: dot positions}, the left dot moves back and forth around its starting position, whereas the right dot generally moves towards the upper right in the x-y plane. From the right dot's relative shift of $\sim$ 36 nm, we estimated the change to $\Delta_{HL}$. Using the simulated data in Ref. \cite{Corley-Wiciak2023}, we note a maximum change of $\sim 4\%$ in the HH-LH splitting. Compared to the device in this study, the quantum well in Ref. \cite{Corley-Wiciak2023} is situated closer to the confinement gates, which greatly increases the effects of strain. Consequently, we only wish to use this 4$\%$ change in $\Delta_{HL}$ as a guide for understanding the pattern of our data. We expect the modification to $\Delta_{HL}$ to be minimized directly underneath an electrode and greatest between two electrodes, which is a distance of $50$ nm in our device. Naively assuming a linear dependence of the strain on the dot's position underneath the gates, we can estimate the change in $\Delta_{HL}$ from the 36 nm shift in the right dot: 

\begin{equation*}
    \delta \Delta_{HL} \sim \frac{4\%}{50 \text{nm}} 36 \text{nm} = 3\%. 
\end{equation*}

\end{document}